\documentstyle[epsfig]{aprim10}
\input{epsf}
\newif\ifAMStwofonts
\title[Dynamics of Neptune's Trojans]{On the Dynamics of Inclined Neptune's Trojans}

\author[Zhou et al.]
       {Li-Yong Zhou$^1$, Rudolf Dvorak$^2$ and Yi-Sui Sun$^1$\\
        $^1$Department of Astronomy, Nanjing University, Nanjing
        210093, China\\
        $^2$Institute of Astronomy, University of Vienna,
        T\"{u}rkenschanzstr. 17, 1180 Vienna, Austria}
\date{}

\pagerange{\pageref{firstpage}--\pageref{lastpage}} \pubyear{2008}

\begin{document}

\maketitle

\begin{abstract}
The dynamics of artificial asteroids on the Trojan-like orbits
around Neptune is investigated in this paper. We describe the
dependence of the orbital stability on the initial semimajor axis
$a$ and inclination $i$ by constructing a dynamical map on the
$(a,i)-$\,plane. Rich details are revealed in the dynamical map,
especially a unstable gap at $i=45^\circ$ is determined and the
mechanism triggering chaos in this region is figured out. Our
investigation can be used to guide the observations.
\end{abstract}


\section{Introduction}
In the restricted three-body model consisting of the Sun, a planet
and an asteroid, the equilateral triangular Lagrange equilibrium
points ($L_4$ and $L_5$) are stable for all planets in our Solar
system. Asteroids in the vicinities of $L_4$ and $L_5$ are called
Trojans after the the group of asteroids found around Jupiter's
Lagrange point. Trojan asteroids of Mars and Earth have also been
observed while the Trojan-type orbits of Saturn and Uranus have
been proved unstable due to the perturbations from other planets.
As for Neptune, 6 Trojan-like asteroids were discovered in recent
years (IAU: Minor Planet Center, http://www.cfa.harvard.edu/
iau/lists/NeptuneTrojans.html). We list their orbital properties
in Table 1. The long-term orbital stability of these asteroids has
been studied and verified in different papers, e.g.
\cite{mar03,bra04b,lij07}. There could be much more Trojan-type
asteroids sharing the orbit with Neptune than with Jupiter, both
in the sense of number and total mass \cite{she06}. Therefore it
is worth to investigate the stable region in the whole parameter
space. This topic can be found in several papers
\cite{nes02a,dvo07}.

Since the $L_5$ point of Neptune is nowadays in the direction of
the Galaxy center and not suitable for asteroid observing, all the
asteroids in Table 1 are around the $L_4$ point. There are reports
that the shape and size of the stable regions around $L_4$ and
$L_5$ points are different from each other \cite{hol93}, but
further analysis prove that this asymmetry is no more than an
artificial effect of asymmetric initial conditions
\cite{nes02a,dvo08}. Thus, it is reasonable to study only one of
the Lagrange points and expect the other one has the same
dynamical behavior.

All the Trojans in Table 1 are on the near-circular orbits (small
eccentricities) and two of them have high inclination values. The
origin of the high-inclined orbit is an interesting topic
\cite{lij07}, but in this paper, we will discuss only the stability
of inclined orbits and try to find out the possible region (in
dynamical sense) where the potential Neptune Trojans could survive
for long time.

\begin{table}
\caption{Orbits of Neptune's Trojans. The mean anomaly $M$ is given
at epoch TD=20080514. The perihelion argument $\omega$, ascending
node $\Omega$ and inclination $i$ are in degree (J2000.0).}
 \center{\begin{tabular}{|l|c|c|r|r|r|r|c|c|}
 \hline
 Designation  & $M$ & $\omega$ & $\Omega$ & $i$ & $e$ & $a$\,(AU)\\
 \hline
 2001 QR322  & 58.89  & 158.5 & 151.6 &  1.3 & 0.031 & 30.262 \\
 2004 UP10   & 339.30 & 359.3 & 34.8  &  1.4 & 0.027 & 30.171 \\
 2005 TN53   & 284.82 &  86.7 &  9.3  & 25.0 & 0.064 & 30.143 \\
 2005 TO74   & 265.46 & 304.0 & 169.4 &  5.3 & 0.052 & 30.151 \\
 2006 RJ103  & 231.55 &  32.7 & 120.8 &  8.2 & 0.027 & 30.036 \\
 2007 VL305  & 351.72 & 215.1 & 188.6 & 28.1 & 0.062 & 30.007 \\
 \hline
 \end{tabular}}
\end{table}

\section{Dynamical map}
To investigate the effects of inclination on the stability of
Trojans, we numerically simulate the evolutions of thousands of test
particles on the Trojan-like orbits. The dynamical system consists
of the Sun, four jovian planets (Jupiter, Saturn, Uranus and
Neptune) and the massless test particles. For each set of initial
conditions, a specific inclination value is given and 101 artificial
Trojans are initialized around the $L_5$ point of Neptune as follow:
Their eccentricities $e_0$, ascending nodes $\Omega_0$ and mean
anomalies $M_0$ are exactly the same as the ones of Neptune, but the
perihelion arguments $\omega_0$ differs $60^\circ$ from the one of
Neptune. Because the Trojans share the same orbit with the planet,
they are in the 1:1 mean motion resonance with the planet. The
critical argument of this resonance is $\sigma=\lambda-\lambda_N$
where $\lambda =\omega+ \Omega+ M$ is the mean longitude and the
subscript `N' denotes Neptune. In this paper, we always study the
case of the trailing Lagrange point ($L_5$), so that $\sigma_0=
-60^\circ$. The semimajor axes $a_0$ of test particles are from
29.9\,AU to 30.5\,AU with an increment 0.006\,AU (the osculating
semimajor axis of Neptune is 30.14\,AU at the starting of
simulation). Finally, we vary their inclinations from $0^\circ$ to
$70^\circ$ with an increment of $1.25^\circ$. The systems are then
integrated to $2.687\times 10^7$\,yrs with a Lie-integrator
\cite{han84}. An on-line low-pass digital filter is applied to
filter the high-frequency terms in the output and reduce the final
data size. We apply spectral analysis to the final data and use the
{\it spectral number} to indicate the regularity of an orbit. In
principle, the spectral number (hereafter SN) is the number of peaks
in a power spectrum which are higher than a specific threshold. For
details of calculating SN, see for example \cite{sfm05}.

Figure\,1 shows the dynamical map on the initial $(a_0,i_0)$
plane. The grey depth indicates the SN of the resonant argument
$\sigma$. The spectral number is forced to be 100 if it is greater
than 100, and those orbits with averaged semimajor axis $\bar{a}
\notin [29.9,30.5]$ (AU) are given a spectral number of 110
(dashed area in Figure\,1) since they are surely not inside the
1:1 resonance. Those orbits with small SNs are dominated by few
dominating frequencies thus are more regular while the bigger SNs
indicate strong noises in the motion and chaotic orbits. To verify
the reliability of this regularity indicator, we also integrate
hundreds of orbits to the solar system age (4.5G years) using the
{\it hybrid symplectic integrator} in the {\it Mercury6}
integrator package \cite{cha99}. The comparison between the SNs
derived from our $2.687\times 10^7$\,yrs integrations and the
results from the 4.5G years integrations done by the Mercury6 show
that orbits with small SNs are generally regular and survive on
the Trojan-like orbits in 4.5G years while orbits with SNs higher
than $\sim 60$ will escape from the resonant region and/or be
ejected by (collide with) the planets or the Sun.

\begin{figure}
 \vspace{6.5cm}
 \includegraphics{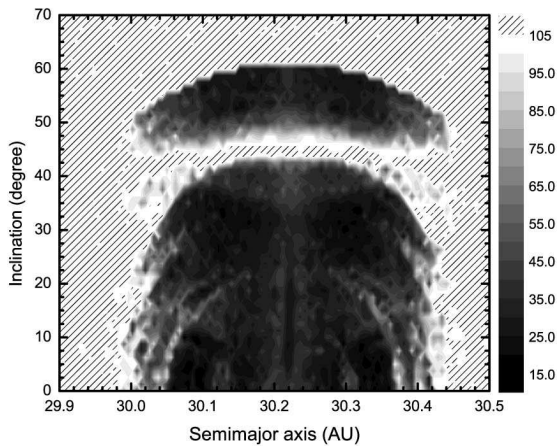}
 \caption{The dynamical map. The spectral number of the resonant argument
 $\sigma$ is mapped on the initial ($a_0,i_0$) plane. }
\end{figure}

There are some interesting features in Figure\,1 deserving a
description.

1) The center of the resonant orbits is at $\sim 30.218$\,AU in
terms of the initial semimajor axis while the mean semimajor axis is
$\sim 30.11$\,AU. The stable region is distributed symmetrically
with respect to this center. Away from the center, the libration
amplitude of the resonant argument $\Delta\sigma$ increases as shown
in Figure\,2 and the Trojans run on the so-called `tadpole' orbit.
It is reported that no Neptune's Trojan can survive with
$\Delta\sigma> 60^\circ$ \cite{hol93,nes02a}. Comparing Figure 1 and
2, we can derive the same conclusion. The inner and outer edges of
the resonant region seen in Figure\,1 $a_0\sim 30.04$\,AU and $\sim
30.40$\,AU are defined by the overlapping of the secondary
resonances.

2) The stable orbits with initial inclination as high as
$i_0=60^\circ$ exist as Figure 1 indicates. This upper limit in fact
can be found in a restricted three-body model in which it's
$61.7^\circ$ \cite{bra04a}. Different values ($35^\circ$ and
$70^\circ$) of such a threshold \cite{nes02a,dvo07} probably are due
to the specific initial conditions of the orbits, the stability
criteria and the inadequate sample Trojans in their simulations.

3) The most distinguishable feature in Figure 1 is a gap locating
at $i_0 \sim 45^\circ$. It separates the resonant region into two
disconnected parts. Therefore we may expect to find two primordial
Trojan groups in future, because this gap prohibits Trojans in one
region from entering the other through secular diffusion.

\begin{figure}
 \vspace{6.50cm}
 \includegraphics{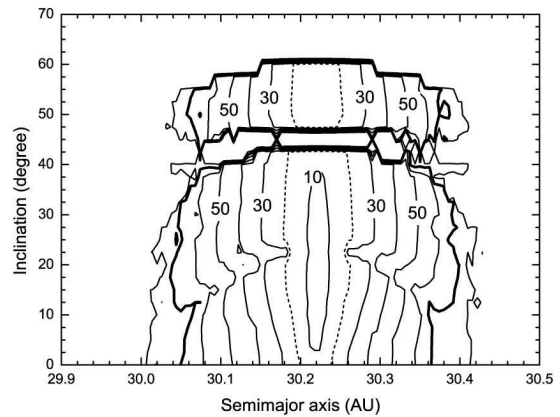}
 \caption{Contour of libration amplitude of the resonant argument. The
 curves of $\Delta\sigma=10^\circ, 30^\circ,50^\circ$ are labeled. The dashed
 and thick solid curves are for $\Delta\sigma=20^\circ$ and $70^\circ$, respectively. }
\end{figure}

4) There are rich details in the resonant region. For example, two
less regular regions at low inclination and with $a_0\sim 30.11$ and
$\sim 30.31$\,AU are visible in Figure\,1. They arise from the
secular resonance $\nu_{18}$, that is,  the nodal frequency of the
Trojan in these regions are very close to Neptune's nodal frequency,
$\dot\Omega \sim \dot\Omega_N$. Another noteworthy structure is an
arc of irregular motion extending from ($a_0=30.04$\,AU,
$i_0=0^\circ$) to ($a_0=30.15$\,AU, $i_0=23^\circ$), and the
symmetrical arc on the right side also. Although the mechanism
behind is not understood very well, the arc is reflected in
Figure\,2 again as a valley of large libration amplitude.

Up to now all the initial eccentricities of test Trojans are set
to be the same as Neptune ($e_0=0.006$). In the dynamical
evolution, the eccentricities of stable orbits are kept small. In
fact for most of stable orbits with $i_0<45^\circ$, the maximum
eccentricity is smaller than 0.05. Only some Trojans with
$i_0>45^\circ$ may be on eccentric orbits, but the eccentricities
are still limited by $e\sim 0.1$.

\section{Motion in the gap}
The most prominent trait in the dynamical map (Figure 1) is the
gap at $i_0\sim45^\circ$. We applied the frequency analysis to
orbits starting from the gap and have figured out the mechanism
causing the chaotic motion. In Figure\,3, we illustrate a typical
orbit in the gap and use this as an example to explain how an
orbit in the gap evolves.

\begin{figure}
 \vspace{11.0cm}
 \includegraphics{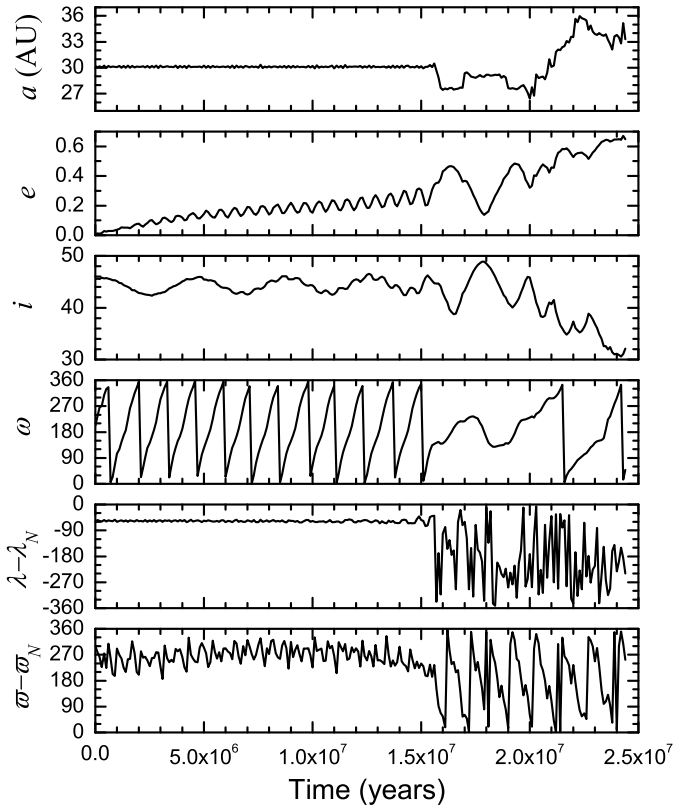}
 \caption{A typical orbit in the unstable gap in Figure\,1. The initial orbital
 elements of this orbit are $a_0=30.218$\,AU, $e_0=0.006, i_0=46.25^\circ$.
 From top to bottom, the panel shows the evolution of the semimajor axis,
 eccentricity, inclination, perihelion argument, resonant argument and the
 difference between the perihelion longitudes of the Trojan and of Neptune.
 The integration of this orbit was terminated at $2.45\times 10^7$\,yrs when
 this Trojan was ejected by a close encounter with Uranus.}
\end{figure}

As shown in Figure\,3, the resonant argument $\sigma= \lambda-
\lambda_N$ librates with a small amplitude around the Lagrange
point $L_5$ with $\sigma\in(-65^\circ, -52^\circ)$ before
$1.57\times 10^7$\,yrs. This libration implies that the Trojan is
in the 1:1 mean motion resonance with Neptune. Thanks to the
protection of the resonance, the evolutions of other orbital
elements during this period are also regular. For example the
semimajor axis is nearly constant and the inclination variation
around $\sim44^\circ$ with an amplitude of only $\sim4^\circ$. But
there is one exception, the eccentricity of the Trojan keeps
increasing during this period and, the eccentricity reaches
$e=0.355$ at $T=1.57\times 10^7$\,yrs. We know that the secular
resonance related to the precession of the perihelion may drive
the eccentricity up \cite{mur99}. We check the frequency of the
perihelion longitudes of the Trojan and the planets in the system,
the result show that the secular resonance $\nu_8$ ($\dot\varpi
\approx \dot\varpi_N$) is responsible for the eccentricity
increasing. A proof of this secular resonance is clearly shown in
the bottom panel in Figure 3 where $\varpi - \varpi_N$ librates
around $\sim 270^\circ$.

This high eccentricity makes the Trojan's perihelion distance
$q=a(1-e)\approx 19.4$\,AU, which means the Trojan may cross the
orbit of Uranus ($a_U=19.2$\,AU). But its high inclination makes the
probability of close encounter with Uranus very small thus the orbit
can be still safe. However, the high eccentricity also makes another
secular resonance possible, the Kozai resonance \cite{koz62}, in
which the perihelion argument $\omega$ librates while the
eccentricity and inclination undergo variations such that the
quantity $H_K=\sqrt{1-e^2}\cos i$ remains constant. These can be
found after $T=1.57\times 10^7$\,yrs in the 2nd, 3rd and 4th panel
of Figure\,3.

In Kozai resonance, when the eccentricity increases the inclination
decreases. Consequently the probability of close encounter with
Uranus or other planets is enhanced significantly, and such close
encounters make the Trojan's orbit unstable, as Figure 3 shows.

\section{Conclusion}
The dependence of the stability of Neptunian Trojans on their
inclinations is investigated and shown by a dynamical map. A gap
of unstable orbits with initial inclination of $\sim 45^\circ$ has
been discovered. The mechanism responsible for this unstable gap,
a combined effects from the $\nu_8$ secular resonance and the
Kozai resonance has been figured out.

\section*{Acknowledgements}
This work was supported by the Natural Science Foundation of China
(No. 10403004, 10833001, 10803003), the National Basic Research
Program of China (2007CB814800). LYZ thanks University of Vienna for
the financial support during his stay in Austria.

\label{lastpage}

\clearpage

\end{document}